\documentclass[aps,prb,reprint,superscriptaddress]{revtex4-1}
\usepackage{graphicx}
\usepackage{dcolumn}

\begin{document}

\title{Two-dimensional ReN$_2$ materials from first principles}

\author{Shi-Hao Zhang}
\affiliation{Beijing National Laboratory for Condensed Matter Physics, Institute of Physics, Chinese Academy of Sciences, Beijing 100190, China}
\author{Bang-Gui Liu}
\email[]{bgliu@iphy.ac.cn}
\affiliation{Beijing National Laboratory for Condensed Matter Physics, Institute of Physics, Chinese Academy of Sciences, Beijing 100190, China}
\affiliation{School of Physical Sciences, University of Chinese Academy of Sciences, Beijing 100190, China}

\date{\today}

\begin{abstract}
The discovery of graphene makes it highly desirable to seek new two-dimensional materials. Through first-principles investigation, we predict two-dimensional materials of ReN$_{2}$: honeycomb and tetragonal structures. The phonon spectra establish the dynamical stability for both of the two structures, and the calculated in-plane stiffness constants proves their mechanical stability. The energy bands near the Fermi level consist of N-p and Re-d orbitals for the honeycomb structure, and are mainly from Re d orbitals for the tetragonal structure. While the tetragonal structure is non-magnetic, the honeycomb structure has N-based ferromagnetism, which will transit to anti-ferromagnetism under 14$\%$ biaxial strain. The calculated electron localization function and spin density indicate that direct N-N bond can occur only in the honeycomb structure. The ferromagnetism allows us to distinguish the two 2D phases easily. The tetragonal phase has lower energy than the honeycomb one, which means that the tetragonal phase is more stable, but the hexagonal phase has much larger bulk, shear, and Young's muduli than the tetragonal phase. The tetragonal phase is a three-bands metal, and the hexagonal phase is a ferromagnetic semi-metal. The special structural, electronic, magnetic, and optical properties in the honeycomb and tetragonal structures make them promising for novel applications.
\end{abstract}


\maketitle


\section{Introduction}

Recent years have witnessed a booming development in two-dimensional materials since the experimental discovery of graphene~\cite{1,2,3,4}. Besides graphene, such two-dimensional materials as monolayer phosphorus and monolayer transition-metal dichalcogenides (TMDC) exhibit many novel physical properties which do not exist in their bulk counterparts. The two-dimensional materials are promising for applications in many fields, such as field-effect transistors~\cite{5,6}, phototransistors~\cite{7,8}, p-n junctions~\cite{9,10}, supercapacitors~\cite{11,12}, and batteries~\cite{13,14}.

The experimental realization of two-dimensional MoN$_2$ material~\cite{15} means that the two-dimensional material field has been extended to the transition-metal dinitrides (TMDN). Very interestingly, the ferromagnetism is found in two-dimensional MoN$_2$, ZrN$_2$, and TcN$_2$, and furthermore two-dimensional YN$_2$ has half-metallic feature~\cite{16}. It is found that the ferromagnetism in the 2D MoN$_2$ will change into antiferromagnetism when appropriate biaxial strain is applied~\cite{17}. The MoN$_2$ is also proved to be an appealing two-dimensional electrode material with high storage capacity~\cite{18}.
On the other hand, the three-dimensional materials of rhenium dinitride ReN$_{2}$ have been synthesized by metathesis reaction under high pressure~\cite{19}, and X-ray Diffraction(XRD) shows the samples have the same structure as the three-dimensional MoS$_{2}$-like honeycomb structure. A first-principles calculation of the stiffness constants indicates that this structure is not mechanically stable and a monoclinic structure was suggested instead~\cite{20}, but a more systematical first-principles investigation shows that the three-dimensional MoN$_2$-like structure is dynamically stable for ReN$_{2}$ and there is a three-dimensional tetragonal structure with lower total energy~\cite{21}. In addition, experiment indicates that the three-dimensional ReN$_2$ may be layered in terms of its compressibility~\cite{19}. Nevertheless, there has been no exploration for two-dimensional materials of ReN$_2$ monolayers.

Here, through first-principles calculation we study the two-dimensional crystal structures of ReN$_2$ and their electronic, mechanical, magnetic, and optical properties for two-dimensional ReN$_2$ materials. We obtain two stable two-dimensional structures, namely honeycomb-like and tetragonal monolayers, whose stability is confirmed by calculated phonon spectra and in-plane stiffness constants. The energy bands, density of states (DOS), magnetic ordering, orbital occupations, and effect of the spin-orbit coupling are investigated. Both of them are metallic, and the 2D hexagonal phase has ferromagnetism based on nitrogen. The strain-dependent magnetic properties and the optical properties of the two-dimensional materials are also studied to explore possible applications. More detailed results will be presented in the following.

\section{Computational methods}

The first-principles calculations are done with the projector-augmented wave (PAW) potential method~\cite{22} as implemented in the Vienna ab initio simulation package software (VASP)~\cite{23}. For all of our spin-polarized and spin-unpolarized computational cases, we take the generalized gradient approximation (GGA), accomplished by Perdew, Burke, and Ernzerhof (PBE)~\cite{24}, for the exchange-correlation functional. The kinetic energy cutoff of the plane waves is set to 600 eV. We choose 15\AA{} as the thickness of vacuum slab for our computational models of the ReN$_2$ monolayers in order to avoid artificial interaction between the two neighboring monolayers. For both optimization and static calculation, the Brillouin zone integration is carried out with a 15$\times$15$\times$1 special $\Gamma$-centered k-point mesh following the convention of Monkhorst-Pack~\cite{25}. All atomic positions are fully optimized with the conjugate gradient optimization until all the Hellmann-Feynman forces on each atom are less than 0.001 eV/\AA{} and the total energy difference between two successive steps is smaller than 10$^{-8}$ eV. Furthermore, phonon dispersion calculation in terms of the density functional perturbation theory, by using the PHONOPY program~\cite{29}, is performed to ensure the structural stability of the monolayers. We take the 4$\times$4$\times$1 supercell for calculating the phonon spectra of the 2D  structures. In order to make further confirmation, band dispersion calculations with Heyd-Scuseria-Ernzerhof (HSE) hybrid functional~\cite{26,27,28} are carried out, with the mixing rate of the HF exchange potential being 0.25. The dielectric functions and other optical properties of the two-dimensional honeycomb and tetragonal structures are calculated in terms of perturbation theory as implemented in VASP~\cite{23}.


\section{results and discussion}

\subsection{Structures and stability}

\begin{figure}[!htbp]
\includegraphics[width=0.48\textwidth]{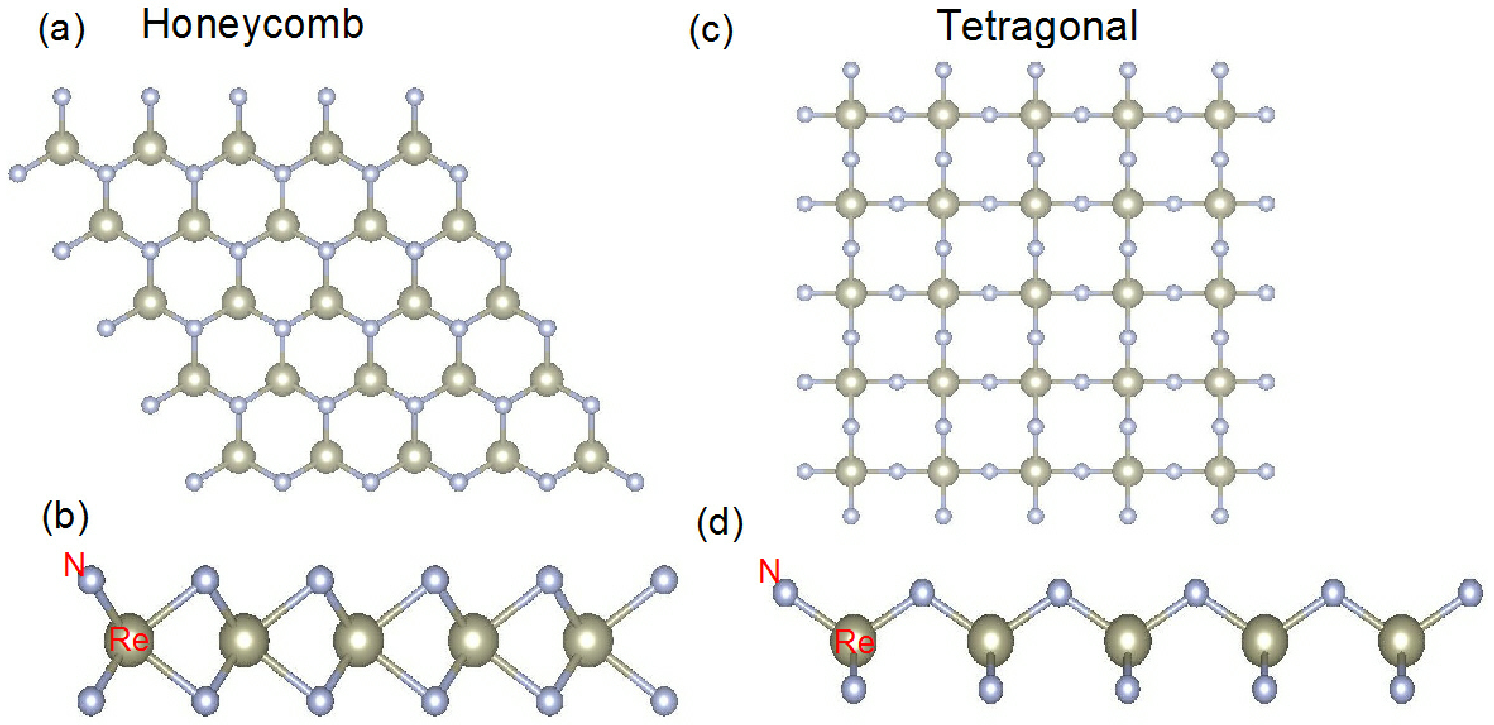}
\caption{\label{fig1}The top and side views of the 2D honeycomb (a,b) and tetragonal (c,d) structures of ReN$_2$.}
\end{figure}

We realize two-dimensional ReN$_{2}$ materials through the 2D honeycomb and tetragonal structures of ReN$_{2}$ monolayer. Other 2D crystal structures have been considered, but are rejected due to lacking structural stability. The 2D honeycomb (P6$_3$/mmc) and tetragonal (P$\bar{4}$m2, \#115) structures are both made up with three atom planes as shown in Fig.~\ref{fig1}. Our computational models are constructed by repeating the 2D monolayers and adding a vacuum layer of 15\AA{} between two neighboring monolayers. There exists a ferromagnetic order in the hexagonal structure, but the tetragonal structure is non-magnetic. The total energy of the tetragonal structure is lower than the hexagonal one by 0.33 eV per formula unit. In Table~\ref{tab:table1} we summarize the lattice parameters $a$, the space groups, the distance $d_{N-N}$ between the top and bottom nitrogen planes, the Re-N bond length $l_{Re-N}$, and the cohesive energy $E_{coh}$. The cohesive energy per formula unit is defined as  $E_{coh}=E_{Re}+E_{N^2}-E_{ReN_2}$, where $E_{Re}$ and $E_{N^2}$ are the total energies of isolated Re atom and N$^2$ molecule. These cohesive energies, 6.11 and 6.44 eV, are much larger than 1.87 eV for the case of MoN$_2$ \cite{mon2ce}.

\begin{table}[!htbp]
\caption{\label{tab:table1}The structural parameters and the cohesive energies of the 2D honeycomb and tetragonal ReN$_2$ materials.}
\begin{ruledtabular}
\begin{tabular}{cccccc}
Structure & $a(\mathrm{\r{A}})$ & Space group & $d_{N-N}(\mathrm{\r{A}})$ & $l_{Re-N}(\mathrm{\r{A}})$ & $E_{coh}$(eV)\\ \hline
Honeycomb & $2.877$ & P6$_3$/mmc & $2.369$ & $2.040$ & $6.11$\\
Tetragonal& 3.178& P$\bar{4}$m2& 1.958& 1.866& 6.44\\
\end{tabular}
\end{ruledtabular}
\end{table}

\begin{figure}[!htbp]
\includegraphics[width=0.51\textwidth]{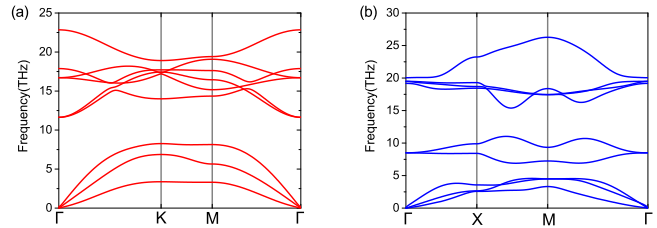}
\caption{\label{fig2}The phonon dispersions of the 2D honeycomb (a) and tetragonal (b) structures of ReN$_2$.}
\end{figure}

In order to analyze the dynamical stability of these two 2D structures, the phonon spectrum calculations are performed by using the PHONOPY program~\cite{29}, which is based on the density functional perturbation theory. The calculated results are presented in Fig.~\ref{fig2}. It is clear that there are nine phonon branches including three acoustic branches and six optical phonon bands. Non-existence of negative phonon frequencies in Fig. 2 proves the dynamical stability of the honeycomb and tetragonal structures. In addition, the T-phase structure, which has been found in other two-dimensional materials~\cite{30}, is proved to be unstable for ReN$_2$ in terms of its phonon spectrum result. Distorted T$^{\prime}$-phase structure~\cite{31,32,33,34} is also proved to be impossible  because our phonon spectrum calculation show that there are very large negative phonon frequencies near the $\Gamma$ point. It can be seen in Fig. 2 that in the vicinity of the $\Gamma$ point, the acoustical branches show linear dispersion, and the out-of-plane acoustical branch appears to be softer than the other two due to the special mode in the two-dimensional materials~\cite{35}.

\begin{table*}
\caption{\label{tab:table2}The mechanical properties of the two 2D ReN$_2$ structures: the elastic constants ($C_{11}$, $C_{12}$, $C_{22}$, $C_{66}$), the bulk modulus $B$, the shear modulus $G$, the Young's modulus $E$, the poisson's ratio $\nu$, and Vickers hardness $HV$.}
\begin{ruledtabular}
\begin{tabular}{cccccccccc}
Structure & $C_{11}$(GPa) & $C_{12}$(GPa) & $C_{22}$(GPa) & $C_{66}$(GPa) & $B$(GPa) & $G$(GPa) & $E$(GPa) & $\nu$ & $HV$(GPa)\\ \hline
Honeycomb & 998.661 & 358.846 & 999.563 & 321.990 & 301.768 & 173.690 & 437.191 & 0.259 & 18.410\\
Tetragonal& 547.722 & 37.929 & 551.143 & 111.636 & 130.525 & 93.056 & 225.564 & 0.212 & 16.091\\
\end{tabular}
\end{ruledtabular}
\end{table*}

In order to further investigate the mechanical stability, we present the calculated elastic constants and other key structural parameters in Table~\ref{tab:table2}. For the two-dimensional materials, only $C_{11}$, $C_{12}$, $C_{16}$, $C_{26}$, $C_{66}$ and $C_{22}$ are meaningful quantities, and the criteria of mechanical stability for monolayer materials require that $C_{11} > C_{12} > 0$ is satisfied. These two 2D materials are mechanically stable because Table~\ref{tab:table2} shows that the in-plane stiffness constants of the two structures satisfy the criteria of mechanical stability.

\begin{figure*}[!htbp]
\includegraphics[width=0.85\textwidth]{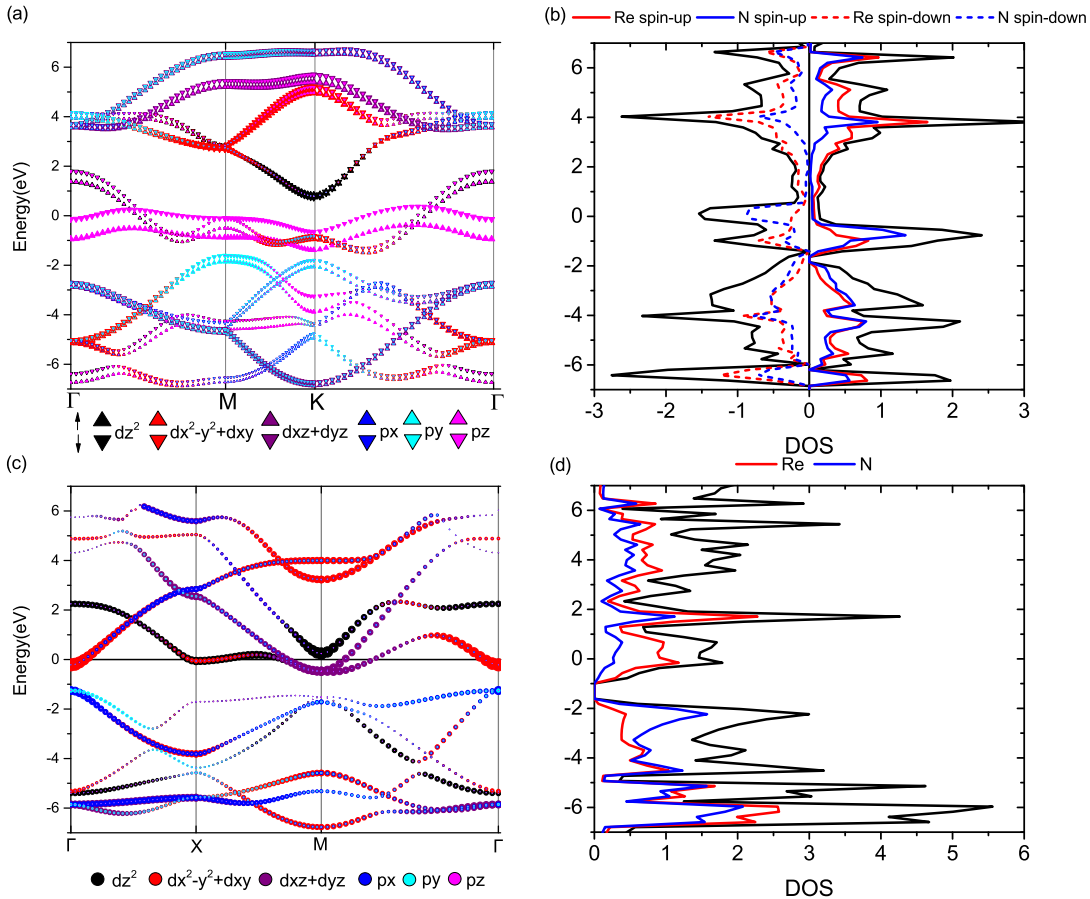}
\caption{\label{fig3}The energy bands and density of states (DOS) of the 2D honeycomb (a,b) and tetragonal (c,d) ReN$_2$ structures. In the band structures, the contribution of different orbits is shown in the form of different colors. The size of the symbol is proportional to the weight of the orbit.}
\end{figure*}

\subsection{Electronic structures}

The 2D honeycomb structure has the $D_{3h}$ symmetric group including the $C_3$ rotation group and mirror inversion $M$, and the 2D tetragonal structure obeys the $S_4$ rotation group. For the electronic wave function of the d orbits, the parity of mirror inversion is even for $d_{z^2}$, $d_{xy}$, and $d_{x^2-y^2}$, and odd for $d_{xz}$ and $d_{yz}$. The band dispersions of these two 2D materials without the spin-orbit coupling, with the weights of orbitals indicated,  are presented in Fig.~\ref{fig3}. The size of symbol is proportional to the weight of the orbital. The band structures show that these two structures are both metallic.

\begin{figure}[!htbp]
\includegraphics[width=0.51\textwidth]{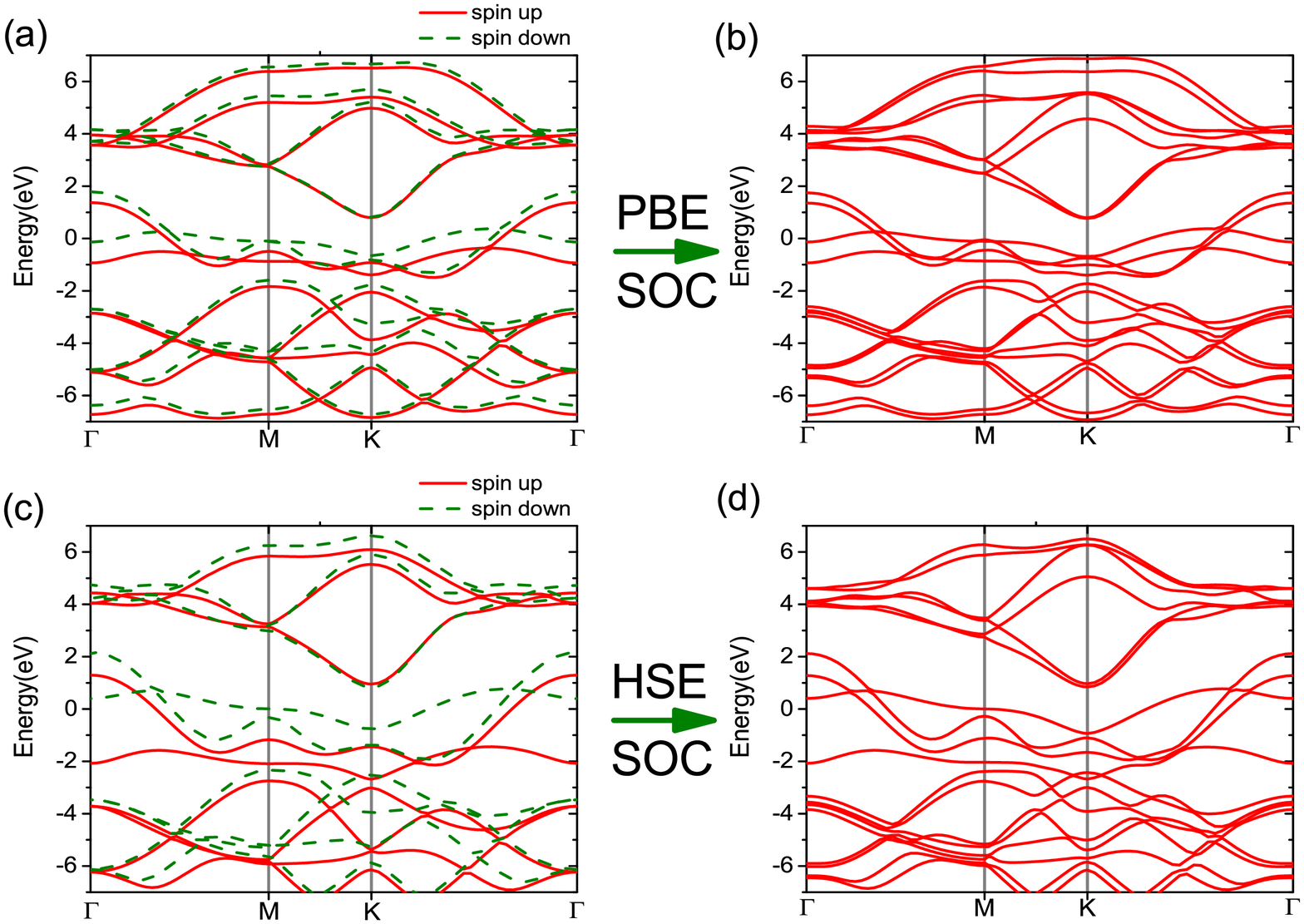}
\caption{\label{fig4}The energy bands of the 2D honeycomb ReN$_2$ structure without (a,c) or with the spin-orbit coupling (b,d). The upper part (a,b) is calculated with the GGA functional, and the lower (c,d) with the HSE functional.}
\end{figure}
\begin{figure}[!htbp]
\includegraphics[width=0.51\textwidth]{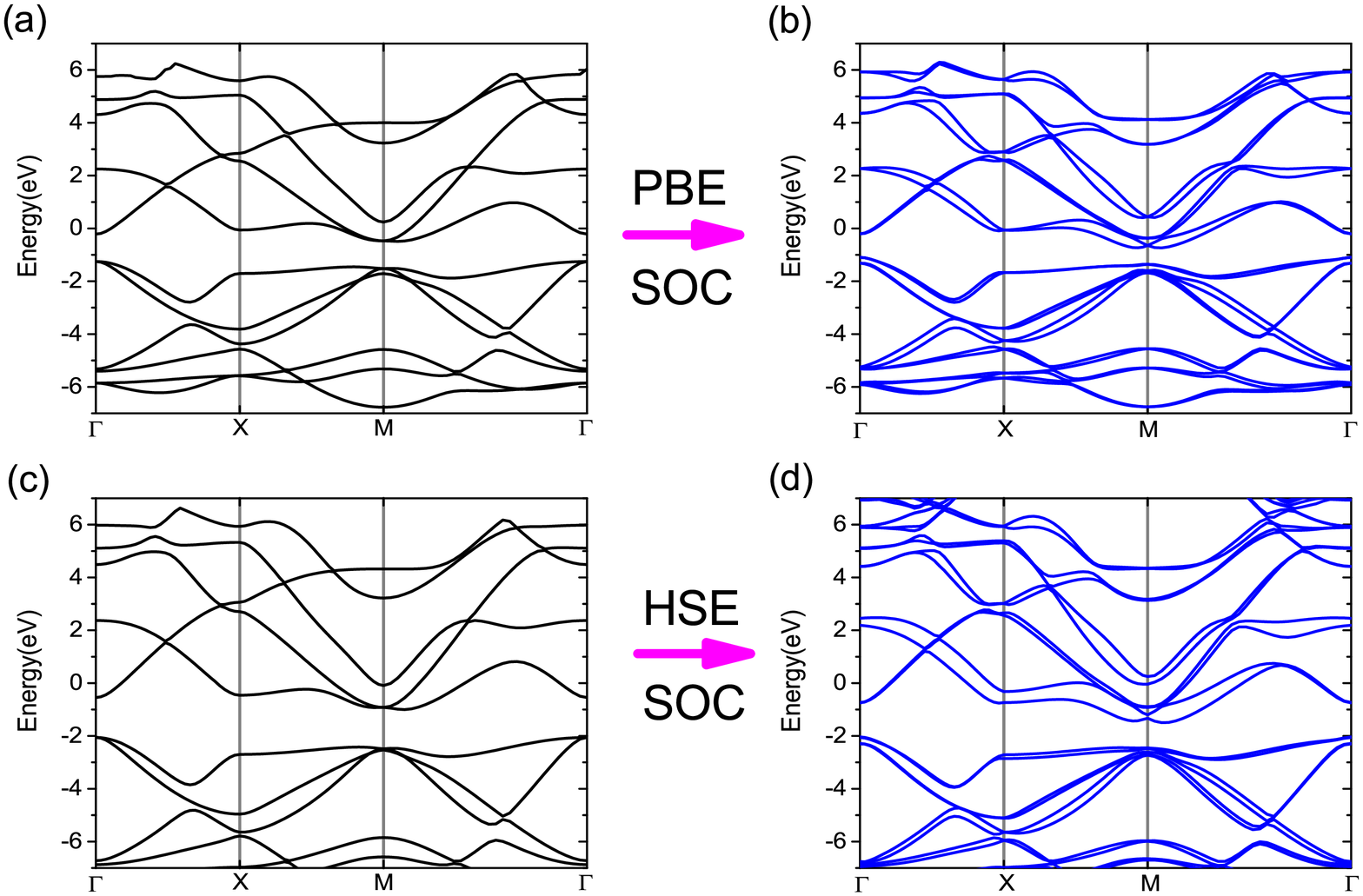}
\caption{\label{fig5}The energy bands of the 2D tetragonal ReN$_2$ structure without (a,c) or with the spin-orbit coupling (b,d). The upper part (a,b) is calculated with the GGA functional, and the lower (c,d) with the HSE functional.}
\end{figure}

For the 2D honeycomb structure, the band structure is spin polarized due to the ferromagnetism. The four energy bands near the Fermi level are mainly from N $p_z$ orbit and Re d$_{z^2}$, d$_{x^2-y^2}$, and d$_{xy}$. Among all the bands, the two $p_z$ bands near the Fermi level are spin polarized most, with the spin-up $p_z$ band being fully filled and the spin-down band partially filled. For the two partially filled d-dominant bands near the Fermi level, the part near $\Gamma$ point is from Re d$_{z^2}$, d$_{x^2-y^2}$, and d$_{xy}$, and the part near the K point originates mainly from Re d$_{x^2-y^2}$. The spin-down DOS peak at the Fermi level is mainly from the spin-down N $p_z$ state. The spin-up peak near -1 eV originates mainly from the spin-up N $p_z$ states. These four bands are decisive for the ferromagnetism and for further electronic transport properties.

As for the the tetragonal phase, the band structure is quite different. It consists of six fully filled bands below -1 eV, three nearly empty bands near the Fermi level, and two empty bands above the Fermi level. In contrast, all the N p bands are filled, and the nearly empty bands come from Re d states. The valence band maximum occurs at the $\Gamma$ point which consists of N $p_x$/$p_y$ orbits and Re $d_{xz}$/$d_{yz}$ orbits. The conduction band minimum locates near the $M$ point, which is made up with Re $d_{xz}$ and $d_{yz}$ orbits. At the $\Gamma$ point, the band near the Fermi level is mainly from Re d$_{x^2-y^2}$, and near the X point, the band near the Fermi level consists mainly of Re d$_{z^2}$ and d$_{x^2-y^2}$. The big difference in the band structures between the hexagonal and tetragonal phases comes from their different crystal structures and symmetries.

Because the rhenium atom's spin-orbit coupling effect should be strong, the energy band calculation with the spin-orbit coupling is necessary. With the spin-orbit coupling the spin is no longer good quantum number because of $[\hat{S},\hat{H}_{soc}]\neq 0$, where $\hat{H}_{soc}$ is the Hamiltonian of the system with the spin-orbit coupling. As a result, it is not necessary to make a distinction between the spin-up and the spin-down in the energy bands. For both of the 2D phases, the energy bands calculated with both GGA and HSE are presented in Fig.~\ref{fig4} and Fig.~\ref{fig5}, respectively, to distinguish the effect of the spin-orbit coupling. The spin-orbit-coupling splitting in the HSE bands is larger than in the GGA bands. For the 2D honeycomb structure, the total energy with the spin-orbit coupling is -28.169 eV per formula unit when the spin direction is perpendicular to the monolayer plane, and the easy magnetic axis is in the plane of monolayer because the in-plane total energy is  0.2 meV lower than the perpendicular direction. As for the tetragonal structure, the total energy with the spin-orbit coupling taken into account, -28.569 eV per formula unit, is approximately 0.33 eV lower than that of the 2D honeycomb phase.

\subsection{Magnetic properties and strain effects}

The 2D honeycomb structure exhibits ferromagnetism while the tetragonal one is nonmagnetic, which is proved by our systematical calculations and comparison of the different magnetic configurations. In addition, the calculation based on Heyd-Scuseria-Ernzerhof (HSE) hybrid functional is also performed for confirming the main results. The HSE functional produces larger spin exchange splitting than the GGA functional.

\begin{figure}[!htbp]
\includegraphics[width=0.5\textwidth]{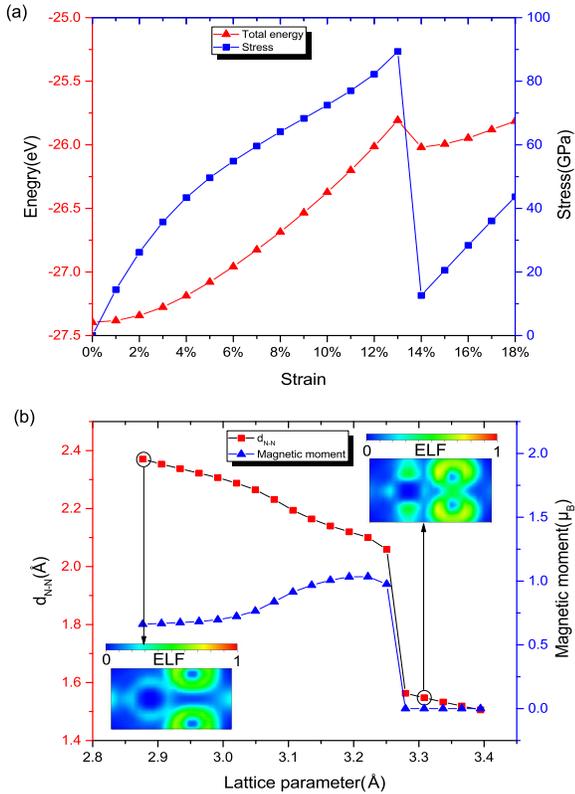}
\caption{\label{fig6} The strain dependence of the total energy and the biaxial stress (a) and the N-N distance $d_{N-N}$ and magnetic moment (b) for the 2D honeycomb structure. The ELF plots  are shown for the zero strain and the 15$\%$ strain.}
\end{figure}

\begin{figure}[!htbp]
\includegraphics[width=0.48\textwidth]{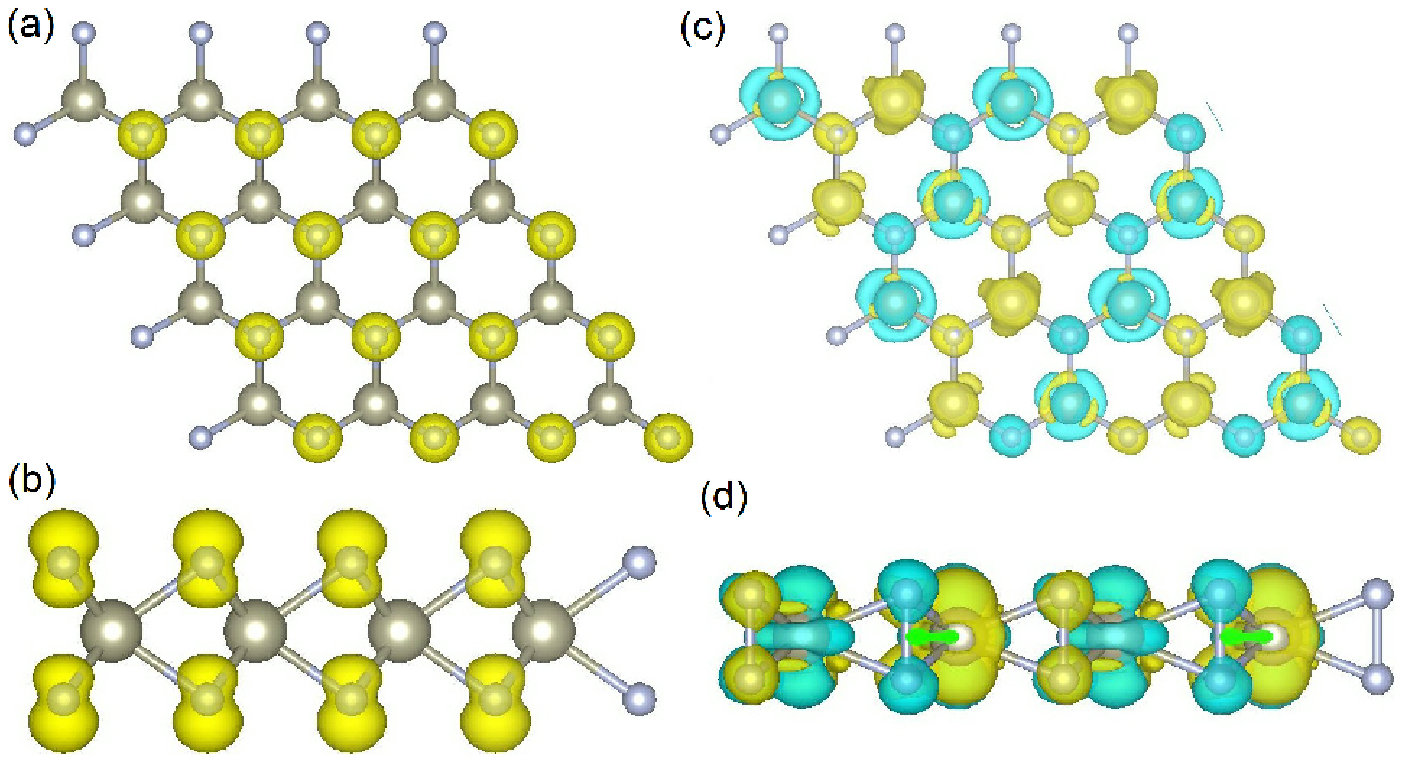}
\caption{\label{fig7}The top (a,c) and side (b,d) views of the spin density in the 2D honeycomb structure under the zero strain (a,b) and 15$\%$ strain (c,d). The yellow denotes the spin-up density and the blue the spin-down density.}
\end{figure}

As shown in Fig.~\ref{fig6} and Fig.~\ref{fig7}, the honeycomb structure has ferromagnetic moment based on nitrogen atoms. The magnetic moment reaches 0.66$\mu_B$ per formula unit, which originates from nitrogen's $p_z$ orbit as shown in Fig.~\ref{fig7}. This implies that one N atom contributes 0.33$\mu_B$ to the magnetic moment. Fig.~\ref{fig3} has already revealed that the largest spin exchange splitting occurs in the N-$p_z$ dominated energy bands. The total energy difference between the antiferromagnetic and ferromagnetic configurations of the honeycomb structure is equivalent to 59 meV, which indicates that the spin exchange interaction is quite strong. Very interestingly, the ferromagnetic honeycomb structure will change when we apply a biaxial stress on it, and it will transit to an antiferromagnetic spin configuration at the critical point. In contrast, there is no phase transition under biaxial strain for the 2D tetragonal structure.

When the biaxial strain is increasing, the distance between two nitrogen layers, $d_{N-N}$, decreases slowly until the critical point. When the strain reaches to 14$\%$, the dive of the distance $d_{N-N}$ reveals the occurrence of the structural phase transition. Our calculation reveals that under this strain, the antiferromagnetic configuration has the lower energy than the ferromagnetic one and the ELF plot shows that the direct bond between the top and bottom nitrogen atoms occurs. There is no such phenomena for the tetragonal structure under biaxial strain.

We have obtained the relation between stress and strain for the two 2D structures under biaxial strain by using the relation, $\sigma = \frac{1}{2V}\frac{\partial E(V,\eta)}{\partial \eta}$,
where $\eta$ is the biaxial strain, $V$ the volume of the unit cell, and $E(V,\eta)$ the total energy of the system. Our calculated result for the hexagonal phase is presented in Fig.~\ref{fig6}. It is found that the stress of about 90GPa is needed for the phase transition of the 2D honeycomb structure.

\begin{figure*}[!htbp]
\includegraphics[width=0.8\textwidth]{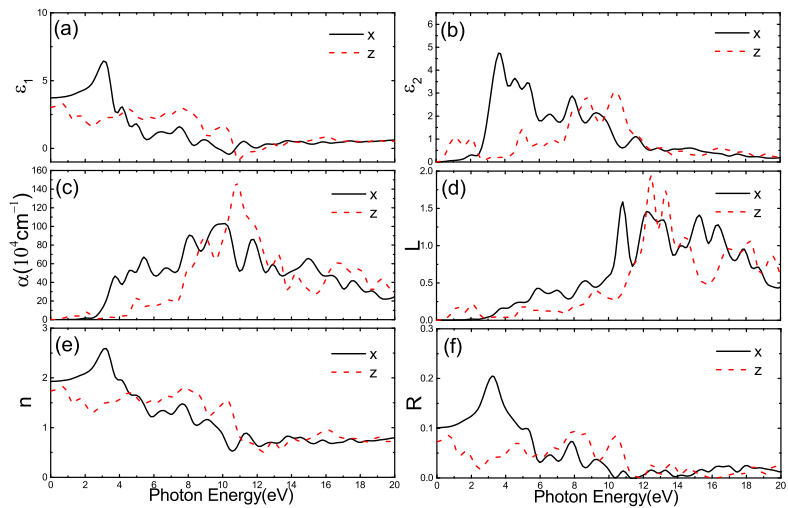}
\caption{\label{fig8} The real part $\varepsilon_{1}$ (a) and the imaginary part $\varepsilon_{2}$ (b) of the dielectric function, the absorption coefficient $\alpha$ (c), the energy loss coefficient $L$ (d), the refractive index $n$ (e), and the reflectance $R$ (f) of the 2D honeycomb ReN$_2$ structure. }
\end{figure*}

\begin{figure*}[!htbp]
\includegraphics[width=0.8\textwidth]{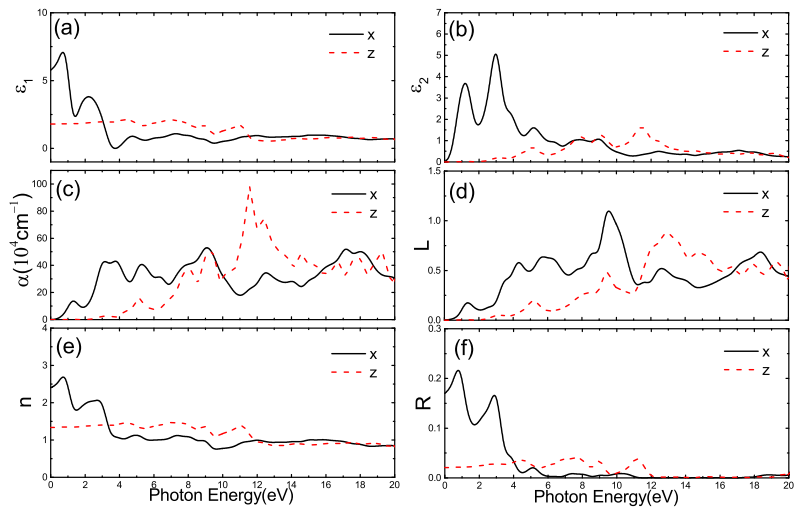}
\caption{\label{fig9}The real part $\varepsilon_{1}$ (a) and the imaginary part $\varepsilon_{2}$ (b) of the dielectric function, the absorption coefficient $\alpha$ (c), the energy loss coefficient $L$ (d), the refractive index $n$ (e), and the reflectance $R$ (f) of the 2D tetragonal ReN$_2$ structure.}
\end{figure*}

\subsection{Optical properties}

The optical properties can be calculated from the dielectric function $\varepsilon$, which can be written as $\varepsilon = \varepsilon_1 +i \varepsilon_2$. The first-principles calculation directly gives the imaginary part of the dielectric function and the real part can be derived by using the well-known Kramers-Kronig relation. Then, we can get the absorption coefficient $\alpha$, energy loss coefficient $L$, refractive index $n$, and reflectance $R$ in terms of the standard relations:
\begin{eqnarray}
\alpha (\omega) = \sqrt{2}\omega\left[ \sqrt{\varepsilon _1^2 +\varepsilon _2^2}-\varepsilon _1 \right]^{1/2},
\end{eqnarray}
\begin{eqnarray}
L(\omega) = \varepsilon _2/(\varepsilon _1^2 +\varepsilon _2^2),
\end{eqnarray}
\begin{eqnarray}
n(\omega) = \left[(\sqrt{\varepsilon _1^2 +\varepsilon _2^2}+\varepsilon _1)/2\right]^{1/2},
\end{eqnarray}
\begin{eqnarray}
R(\omega) = \left|\frac{\sqrt{{\varepsilon}}-1}{\sqrt{{\varepsilon}}+1}\right|^2.
\end{eqnarray}

Because of the isotropic cell for both of the ReN$_2$ monolayers, the in-plane dielectric functions, $\varepsilon _{xx}$ and $\varepsilon _{yy}$, are the same. As a result, the optical functions are isotropic in the xy plane. In contrast, the out-of-plane results are very different from the in-plane ones. We present our in-plane and out-of-plane optical functions in Figs.~\ref{fig8} and \ref{fig9}. It is clear that there is a large difference between the in-plane and out-of-plane curves in each of the optical functions. These reflect the two-dimensional features in the electronic structures for the 2D phases.

In the visible light region ($\omega\sim 1.5-3$eV), the absorption coefficient of the honeycomb structure is almost zero for $\omega< 2.5$eV, while the tetragonal structure's absorption coefficient ranges from 8$\times 10^{4}$cm$^{-1}$ to 46 $\times 10^{4}$cm$^{-1}$, which shows quite good absorption for photovoltaic applications. The energy loss spectrum reveals the collective excitations, and the existence of many peaks for the honeycomb structure indicates the existence of plasmon resonances. It is noted that there is a one-to-one relationship between the reflectance and the photon energy in the visible region for the honeycomb structure. Thus the honeycomb structure can be used to detect the light detection applications.

\section{conclusion}

In conclusion, we have predicted the two 2D structures of ReN$_2$ monolayer as new members of two-dimensional materials by using first-principles calculations. Their structural stability has been established by their DFT phonon spectra and mechanical properties. The 2D hexagonal phase exhibits ferromagnetism that is robust (with a magnetic energy of 59 meV) against magnetic fluctuations, and can undergo a phase transition under biaxial strain. The stress-driven transition also changes the ferromagnetism to antiferromagnetism. There is an easy magnetic plane in the honeycomb monolayer plane, which implies that the magnetism is like the well-known 2D Heisenberg xy spin model which can host interesting topological spin structures. The tetragonal phase is 0.33 eV (per formula unit) lower than the hexagonal phase, but the hexagonal phase has much larger bulk, shear, and Young's muduli than the tetragonal phase. For the tetragonal phase, the N p bands are fully filled and therefore the electronic states near the Fermi level originate from Re d orbitals only. Consequently, the tetragonal phase is a three-bands typical metal, and the hexagonal phase is a ferromagnetic semi-metal. Optical properties have been investigated for both of the phases. We hope that our study on these polymorphic structures of ReN$_2$ can help realize the 2D ReN$_2$ materials experimentally. With these attractive features, they could be used for electronic, spintronic, and optoelectronic applications.

\begin{acknowledgments}
This work is supported by the Nature Science Foundation of China (No. 11574366), by the Strategic Priority Research Program of the Chinese Academy of Sciences (Grant No.XDB07000000), and by the Department of Science and Technology of China (Grant No. 2016YFA0300701).
\end{acknowledgments}



\end{document}